\documentclass[onecollarge]{svjour2}       
\smartqed  
\usepackage{graphicx}
\usepackage{mathptmx}      
\usepackage{amsmath}
\newcommand{\rmc}{\mathrm{c}}
\newcommand{\rmd}{\mathrm{d}}

\newcommand{\rmS}{\mathrm{S}}
\newcommand{\rmt}{\mathrm{t}}
\newcommand{\rmv}{\mathrm{v}}
\newcommand{\ovw}{\text{ov:w}}
\newcommand{\ovs}{\text{ov:s}}
\newcommand{\spc}[1]{SPC$_{#1}$}
\journalname{J Stat Phys}
\begin{document}

\title{Equation of State for Parallel Rigid Spherocylinders
}


\author{Masashi Torikai}


\institute{M. Torikai \at
Department of Physics Engineering, Faculty of Engineering, Mie University\\
Tel.: +81-59-231-9695\\
Fax:  +81-59-231-9726\\
\email{torikai@phen.mie-u.ac.jp}%
}

\date{Received: date / Accepted: date}

\maketitle

\begin{abstract}
 The pair distribution function of monodisperse rigid spherocylinders is calculated by Shinomoto's method, which was originally proposed for hard spheres.
 The equation of state is derived by two different routes: Shinomoto's original route, in which a hard wall is introduced to estimate the pressure exerted on it, and the virial route.
 The pressure from Shinomoto's original route is valid only when the length-to-width ratio is less than or equal to $0.25$ (i.e., when the spherocylinders are nearly spherical).
 The virial equation of state is shown to agree very well with the results of numerical simulations of spherocylinders with length-to-width ratio greater than or equal to 2.
\end{abstract}

\section{Introduction}\label{sec:introduction}
Predicting the equation of state is a simple approach to probe the validity of approximation methods in the theory of liquids.
For hard-sphere (HS) fluids, several methods have been proposed to derive the equation of state, such as the energy, virial, or compressibility equation of state in terms of an approximate pair distribution function; and the virial expansion of the equation of state~\cite{SimpleLiquids}.
Among the many derivations, the method proposed by Shinomoto is unique.

Shinomoto derived the equation of state for HS fluids by calculating the density profile and pair distribution function~\cite{Shinomoto1983}.
The calculation is based on estimating the minimum work required to overcome the depletion force~\cite{Asakura54} between (i) an HS and a hard wall and (ii) between two HSs.
Shinomoto's method is simpler and more concise than the numerous methods based on integral equations and is as accurate as the semiempirical Carnahan-Stirling equation of state~\cite{SimpleLiquids}.
Using Shinomoto's concept, the accuracy of the pair distribution function for an HS fluid turns out to be midway between that obtained by the Born-Green-Yvon theory and by the Percus-Yevick theory~\cite{Wehner1986}.

Shinomoto's method is not restricted to spherical molecules but is applicable to complex molecules as well.
In this paper, I apply Shinomoto's concept to parallel hard-spherocylinder (SPC) fluids.
By calculating the pair distribution function of the SPC fluid via Shinomoto's method, I obtain the fluid pressure by both Shinomoto's original route and the virial route~\cite{Gray1984}.
The results agree well with numerical simulation data via the former route when the length-to-width ratio is less than and equal to 0.25, and via the latter route for SPCs with length-to-width ratio greater than and equal to 2 up to $\infty$.

\section{Theory}
\label{sec:theory}
\subsection{Shinomoto's Method for Hard-SPC Fluids}
\label{subsec:theory-1}
In this section, I extend Shinomoto's method for a uniform monodisperse fluid consisting of parallel rigid SPCs.
Although the discussion here is restricted to hard SPC fluids, it is also applicable, with slight modifications, to general hard-body fluids.

\begin{figure}[htbp]
 \begin{center}
  \includegraphics[clip]{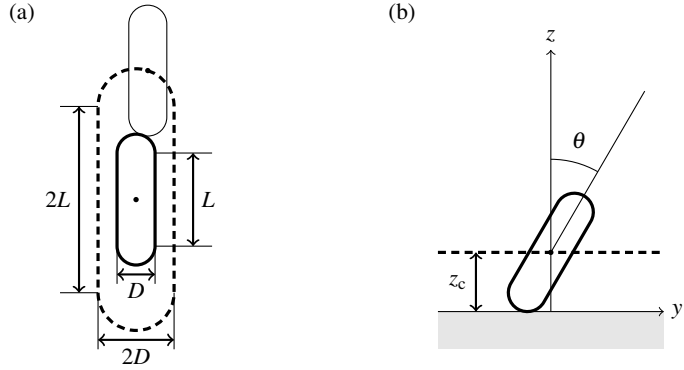}
 \end{center}
 \caption{\label{fig:1}Schematic used to define excluded volume. (a) Each SPC (thick solid curve) has an excluded volume (inside dashed curve) into which the center of another SPC (thin solid curve) cannot enter. (b) Excluded volume (below the dashed line) of hard wall (shaded area).}
\end{figure}

The length and width of the SPCs are represented by $L$ and $D$, respectively, and the number density of the bulk SPC fluid is given by $\rho$.
The packing fraction is denoted by $\eta = \rho v$, with the volume of an SPC being $v=\pi D^{3}/6 + \pi D^{2}L/4$.
The symmetry axes of all SPCs in the fluid are aligned in a common direction.
Since all SPCs are identical and parallel, the excluded volume of an SPC is an SPC with length $2L$ and width $2D$(Fig.\ref{fig:1}a).

The SPC pair distribution function is
\begin{equation}
 g(\vec{r}) = \exp[-\beta \Phi(\vec{r})],
\end{equation}
where $\beta$ is the inverse temperature and $\Phi(\vec{r})$ is the minimum work required to move a test particle (denoted by \spc{\rmt}) from infinity to point $\vec{r}$ under the condition in which another SPC (denoted by \spc{0}) is fixed at the origin of the coordinate system.
The minimum work $\Phi(\vec{r})$ is nonzero because \spc{\rmt} is subjected to a nonvanishing effective force $\vec{F}$ since the density $\rho g(\vec{r})$ is now inhomogeneous because of the existence of \spc{0}.
For example, when \spc{0} and \spc{\rmt} are close enough that their excluded volumes overlap, the pressure exerted on the surface of \spc{\rmt} from surrounding SPCs vanishes inside the excluded volume of \spc{0} because of the absence of any SPC within this volume.
However, outside this excluded volume, the pressure is nonzero.
This mechanism to generate an effective force is similar to that required for the depletion force between particles suspended in the solutions of macromolecules~\cite{Asakura54}.

Let $S(\vec{r})$ be the surface of the excluded volume of \spc{\rmt} at $\vec{r}$, and let $\vec{r}^{*}$ be a point on $S(\vec{r})$.
Since the kinetic pressure at $\vec{r}^{*}$ is $\rho g(\vec{r}^{*})/\beta$, the net force exerted on \spc{\rmt} at $\vec{r}$ is
\begin{equation}
 \vec{F}(\vec{r}) = \frac{\eta}{\beta v} \int_{S(\vec{r})} g(\vec{r}^{*}) \vec{n}(\vec{r}^{*}) \rmd A(\vec{r}^{*}),
\end{equation}
where $\vec{n}(\vec{r}^{*})$ is the inward unit vector normal to $S$ at $\vec{r}^{*}$ and $\rmd A(\vec{r}^{*})$ is the infinitesimal area on $S$ at $\vec{r}^{*}$.
The integral is performed over $S(\vec{r})$.
The minimum work done against $\vec{F}$ is
\begin{equation}
 \Phi(\vec{r}) = -\int_{\infty}^{\vec{r}} \rmd \vec{r}'\cdot\vec{F}(\vec{r}'),
\end{equation}
so the pair distribution function obeys the integral equation
\begin{equation}
 g(\vec{r}) = \exp \biggl[\frac{\eta}{v}  \int_{\infty}^{\vec{r}} \rmd \vec{r}'\cdot\int_{S(\vec{r}')} g(\vec{r}^{*}) \vec{n}(\vec{r}^{*}) \rmd A(\vec{r}^{*})\biggr]. \label{eq:integralEquation}
\end{equation}

In principle, an approximate solution for this integral equation can be derived via an iteration method, as done in Ref.~\cite{Wehner1986} for an HS fluid.
In the present study, however, I follow the original paper by Shinomoto~\cite{Shinomoto1983} and use the result of the first iteration.
As an initial approximation to $g(\vec{r})$, I adopt the low-density limit of the pair distribution function (i.e., $\exp[-\beta u(\vec{r})]$, where $u(\vec{r})$ denotes an intermolecular interaction between molecules separated by $\vec{r}$).
Since the low-density limit of $g(\vec{r})$ is unity outside the excluded volume of \spc{0} and zero inside it, the integration over $S$ on the right-hand side of Eq.\eqref{eq:integralEquation} is zero if the \spc{\rmt} excluded volume does not overlap with that of \spc{0}, and nonzero if they do overlap.
A simple calculation shows that the double integral on the right-hand side of Eq.\eqref{eq:integralEquation} gives the overlap volume of the excluded volumes of \spc{\rmt} and \spc{0}.
Thus,
\begin{equation}
 g(\vec{r}) = \exp[\eta V_{\ovs}(\vec{r})/v], \label{eq:pairDistribution}
\end{equation}
where $V_{\ovs}(\vec{r})$ is the overlap volume of the excluded volumes of two SPCs separated by $\vec{r}$.
Equation \eqref{eq:pairDistribution} is alternatively expressed as $\exp[\eta \int \rmd\vec{r}_{i} f_{0i}f_{i\rmt}]$ in terms of Mayer's $f$-function $f_{0i}$ between \spc{0} and an SPC labeled $i$, and $f_{i\rmt}$ between $i$ and \spc{\rmt}.
The expansion of Eq.\eqref{eq:pairDistribution} in powers of $\eta$ gives an $f$-bond expansion of $g(\vec{r})$ which is exact up to the first order in $\eta$.

To calculate the equation of state as the pressure exerted on a wall, I introduce a hard wall at $z=0$.
Let the angle between the SPC symmetry axes and the $z$ axis be $\theta$, as shown in Fig.\ref{fig:1}b.
An SPC center cannot enter the excluded volume of the wall; that is, the SPC center cannot enter the region $z < z_{\rmc}$, where $z_{\rmc} = (D + L\cos\theta)/2$ is the $z$ coordinate of the point of contact.
To derive the first approximation of the density profile, consider the minimum work to move \spc{\rmt} toward the hard wall (a derivation similar to that of Eq.\eqref{eq:pairDistribution}), which gives
\begin{equation}
 \rho_{1}(\vec{r}, \theta) = \rho_{1}(z, \theta) = \frac{\eta}{v}\exp[\eta V_{\ovw}(z, \theta)/v], \label{eq:rho1}
\end{equation}
where $V_{\ovw}(z, \theta)$ is the overlap volume of the excluded volumes of the wall and SPC at $\vec{r}=(x, y, z)$.

The pair distribution function \eqref{eq:pairDistribution} and the density approximation \eqref{eq:rho1} improve the approximation of the pressure on the test particle and thus the minimum work.
Since the density at point $\vec{r}^{*}$ on $S(\vec{r})$ is $\rho_{1}(\vec{r}^{*}, \theta)g(\vec{r}^{*} - \vec{r})$, the pressure at $\vec{r}^{*}$ takes the form
\begin{equation}
 p_{1}(\vec{r}^{*}, \vec{r}, \theta)
  =
  \rho_{1}(\vec{r}^{*}, \theta)g(\vec{r}^{*} - \vec{r})/\beta
  =
  \frac{\eta}{\beta v} \exp\{\eta [V_{\ovs}(\vec{r}^{*} - \vec{r}) + V_{\ovw}(z^{*}, \theta)]/v\}. \label{eq:p1}
\end{equation}
The minimum work to move a test SPC from infinity to the point of contact with the wall is
\begin{equation}
 \Phi_{\rmc}(\theta) =
  -\int_{\infty}^{z_{\rmc}} \rmd z' \int_{S'(\vec{r})}
  p_{1}(\vec{r}^{*}, \vec{r}', \theta)
  n_{z}(\vec{r}^{*}) \rmd A(\vec{r}^{*}), \label{eq:minimumWork_contact}
\end{equation}
where $S'(\vec{r})$ denotes the subset of $S(\vec{r})$ outside the excluded volume of the wall.
Again, the minimum work \eqref{eq:minimumWork_contact} defines the contact density at the wall, which is $\rho \exp[-\beta \Phi_{\rmc}]$.
Applying the exact pressure sum rule~\cite{Holyst1989} leads to the conclusion that the dimensionless bulk pressure $P^{*}=\beta P v$ is equivalent to the contact packing fraction at the wall, so that the resulting dimensionless pressure is
\begin{equation}
 P^{*}_{\rmS}(\theta) = \eta \exp[-\beta \Phi_{\rmc}(\theta)]. \label{eq:pressure_shinomoto}
\end{equation}
The subscript $\rmS$ indicates that, when using Eq.\eqref{eq:work_contact} to estimate $\beta \Phi_{\rmc}(\theta)$, the pressure is the result of a straightforward extension of Shinomoto's method to parallel SPCs.

By expanding $p_{1}(\vec{r}^{*}, \vec{r}', \theta)$ in powers of the packing fraction $\eta$ up to the second order and substituting the result into Eq.\eqref{eq:minimumWork_contact}, we obtain the following power expansion of $\beta \Phi_{\rmc}(\theta)$:
\begin{equation}
 \beta \Phi_{\rmc}(\theta)
  =
  -4\eta
  -\eta^{2}\frac{1}{v^{2}} \int_{\infty}^{z_{\rmc}}\rmd z' \int_{S'(\vec{r})} [V_{\ovs}(\vec{r}^{*} - \vec{r}') + V_{\ovw}(z^{*}, \theta)] n_{z}(\vec{r}^{*})\rmd A(\vec{r}^{*}). \label{eq:work_contact}
\end{equation}
The coefficient of the first term is independent of $L$ and $\theta$ since it is given by $-V_{\ovw}(z_{\rmc}, \theta)/v$, where $V_{\ovw}(z_{\rmc}, \theta)$ is half of an SPC excluded volume for any $\theta$.
Note that the coefficient of the first term in Eq.\eqref{eq:work_contact} gives the exact second virial coefficient~\cite{Gray2011} if Eq.\eqref{eq:pressure_shinomoto} is expanded in powers of $\eta$.
Since Eq.\eqref{eq:work_contact} contains $V_{\ovw}$, the pair distribution function $\exp[\beta \Phi_{\rmc}]$ cannot be compared to the usual $f$-bond expansion of $g(\vec{r})$, as done below Eq.\eqref{eq:pairDistribution}.

It is possible to obtain $V_{\ovs}(\vec{r})$ and $V_{\ovw}(z, \theta)$ as combinations of the overlap volumes of hemispheres, cylinders, and a half-infinite region.
Here I omit explicit expressions for these functions since the calculation is elementary but the resulting expressions are considerably tedious.

The second term in the surface integral in Eq.\eqref{eq:work_contact} can be reduced to integration over the single variable $z^{*}$, since $V_{\ovw}(z^{*}, \theta)$ is constant when $z^{*}$ is fixed, so the integration over the other variable can be performed analytically.
Similarly, integration of the first term in the surface integral can be reduced to a single variable integration along the SPC axis.
The remaining double integral in Eq.\eqref{eq:work_contact} was calculated numerically.

\subsection{Virial Equation of State}
\label{subsec:theory-2}
In this section, I derive the equation of state through the virial route~\cite{Gray1984} (i.e., not strictly following Shinomoto's method as in the previous subsection) by using the pair distribution function (Eq.\eqref{eq:pairDistribution}) without introducing a hard wall.
The virial equation of state is
\begin{equation}
 P^{*} = \eta - \frac{\beta \eta^{2}}{6v} \int g(\vec{r})\vec{r}\cdot\nabla u(\vec{r}) \rmd^{3} \vec{r}, \label{eq:virialEOS}
\end{equation}
where $u(\vec{r})$ denotes an intermolecular interaction between molecules separated by $\vec{r}$.
In our case, $\nabla u(\vec{r})$ vanishes unless $\vec{r}$ corresponds to a contact configuration, so only the contact value of $g(\vec{r})$ contributes to the virial pressure.

In the molecular frame, in which the $z$ axis lies along the SPC symmetry axis, the contact value of $g(\vec{r})=g(x, y, z)$ is a function of the single variable $z$.
This may be seen by noting that, although $g(\vec{r})$ is a function of $z$ and $x^{2} + y^{2}$, the sum $x^{2} + y^{2}$ is defined by $z$ provided that $\vec{r}$ corresponds to a contact configuration.
The contact value of $g(\vec{r})$ in Eq.\eqref{eq:pairDistribution} is denoted by $g_{\rmc}(z)$ .
The dimensionless virial pressure thus reduces to
\begin{equation}
 P^{*}_{\rmv} = \eta + \frac{2\pi \eta^{2}D^{3}}{3v}
  \biggl[
  \int_{0}^{L} g_{\rmc}(z) \frac{\rmd z}{D} + \int_{L}^{L+1} \biggl\{1 + \frac{(z - L)L}{D^{2}}\biggr\}g_{\rmc}(z) \frac{\rmd z}{D}
  \biggr]. \label{eq:virialEOS2}
\end{equation}

\section{Results}
With isobaric Monte Carlo (MC) simulations~\cite{Frenkel2002}, I obtained ``exact'' results for $g_{\rmc}(z)$ and the equation of state, which are to be compared below with Shinomoto and virial analytical results.
Note that the following MC results are not used in the evaluations of $P^{*}_{\rmS}$ and $P^{*}_{\rmv}$ in the previous section.
In the MC simulations, I used a length-to-width ratio $L/D=5$ and $N=4096$ SPCs.
The SPCs are perfectly aligned along the $z$ axis in a simulation box, and the periodic boundary conditions are imposed in all directions.
I performed $10^{6}$ MC steps and recorded $g_{\rmc}(z)$ and the packing fraction every $10^{3}$ MC steps to calculate their average values.
For isobaric MC simulations, I also checked the consistency of the given pressure by comparing it to the pressure calculated with the right-hand side of Eq.\eqref{eq:virialEOS2} with the pair distribution function calculated via an isochoric MC simulation (the pressures agree within the error bars).
For $L/D=5$ parallel SPCs, MC simulations~\cite{Stroobants1987} give a pressure and the packing fraction at the nematic-smectic phase transition of $P^{*} = 2.11$ and $\eta = 0.406$, respectively, for a system of $N=270$ SPCs.
The results of molecular dynamics simulations~\cite{Veerman1991} are the same for a system of $N=1080$ SPCs.
Since the discussion in the previous section is valid only for the nematic phase, I performed MC simulations at $P^{*} = 0.5$, $1.0$, $1.5$, and $2.0$; under these conditions, no distinct layer structure was found.

\begin{figure}[htbp]
 \begin{center}
  \includegraphics[clip]{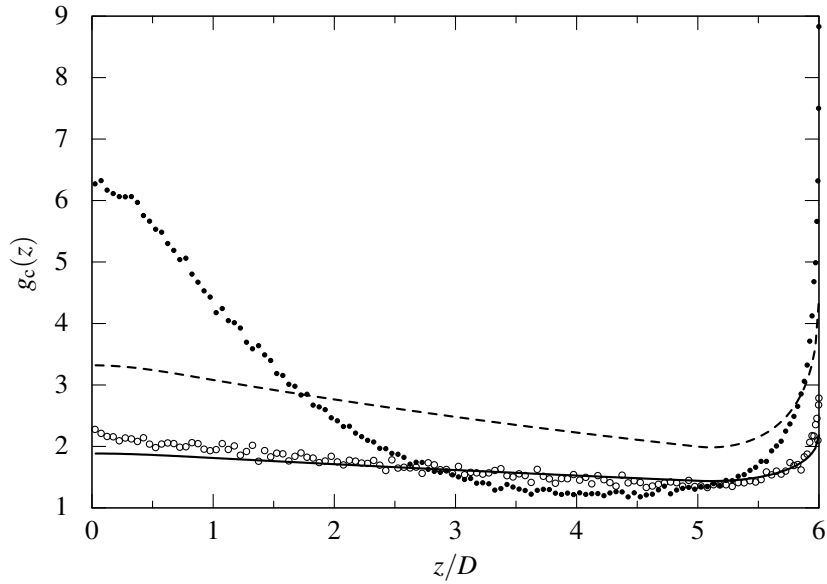}
 \end{center}
 \caption{\label{fig:fig_gc}Contact value of pair distribution function versus $z/D$ (the normalized SPC separation in the symmetry axis direction) for SPCs with $L/D=5$.
 The open and solid circles correspond to isochoric MC simulations for packing fractions $\eta = 0.2076$ (for $P^{*}=0.5$) and $\eta = 0.3929$ (for $P^{*}=2.0$), respectively.
 The solid and dashed curves indicate the respective contact values from Eq.\eqref{eq:pairDistribution} for the same packing fractions.}
\end{figure}

Figure \ref{fig:fig_gc} shows the contact values of $g_{\rmc}(z)$ given by Eq.\eqref{eq:pairDistribution} and those obtained from the MC simulations.
At low density ($\eta=0.2076$), Eq.\eqref{eq:pairDistribution} agrees well with the contact value of the pair distribution function obtained from the isochoric MC simulation.
However, the results diverge at high density.
For example, for $2.8 < z/D < 5$, $g_{\rmc}(z)$ for $\eta = 0.3929$ is lower than $g_{\rmc}(z)$ for $\eta = 0.2076$;
such an inverse dependency on $\eta$ cannot be expected on the basis of Eq.\eqref{eq:pairDistribution} since $g_{\rmc}(z)$ in Eq.\eqref{eq:pairDistribution} increases monotonically as a function of $\eta$ for a given $z$.
A similar failure of Shinomoto's pair distribution function at high density has also been reported for rigid parallel cylinders~\cite{Koda2011}.

\begin{figure}[htbp]
 \begin{center}
  \includegraphics[clip]{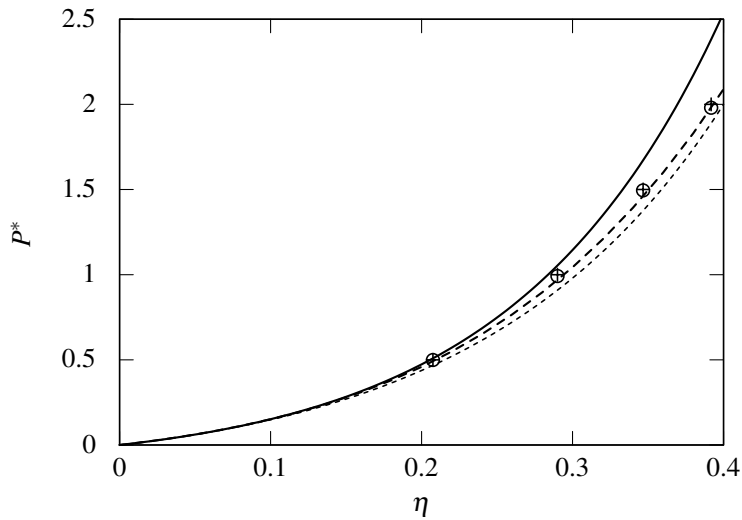}
 \end{center} 
 \caption{\label{fig:eos} Equations of state giving the pressure from Shinomoto's original route and the virial pressure as functions of packing fraction $\eta$ for the nematic phase of parallel SPCs with $L/D=5$.
 The solid and dashed curves indicate $P^{*}_{\rmS}$ and $P^{*}_{\rmv}$, respectively.
 The crosses and circles indicate, respectively, data from isobaric and isochoric MC simulations.
 Their error bars are smaller than the symbol size.
 The dotted curve is the empirically modified result of the scaled particle theory in Ref.\cite{Koda2002}.}
\end{figure}

The agreement between the equation of state and the simulation data is all the more striking given the poor accuracy of our estimate for the pair distribution function at high density.
Figure \ref{fig:eos} shows the equation of state for SPCs with $L/D=5$.
The virial pressure $P^{*}_{\rmv}$ agrees very well with the results of MC simulations throughout the nematic phase.
The empirically modified result of the scaled particle theory for parallel hard SPCs~\cite{Koda2002}, $P_{\text{SPT}}^{*}=(2\eta^{2} + \eta)/(1 - \eta)^{2}$, is also shown in Fig.\ref{fig:eos}.
$P^{*}_{\rmv}$ agrees better with MC simulation results than $P^{*}_\text{SPT}$.

The numerical evaluation of Eq.\eqref{eq:work_contact} indicates that $\Phi_{\rmc}(\theta)$, and thus the pressure $P^{*}_{\rmS}(\theta)$, are independent of $\theta$.
Although this result does not constitute a rigorous proof of the independence of these quantities with respect to $\theta$, it does mean that the $\theta$-dependence of these quantities can be safely neglected.
The pressure $P^{*}_{\rmS}$, which agrees well with the exact result for HSs~\cite{Shinomoto1983}, deviates from the results of MC simulations at high density for SPCs with $L/D=5$.

\begin{figure}[htbp]
 \begin{center}
  \includegraphics[clip]{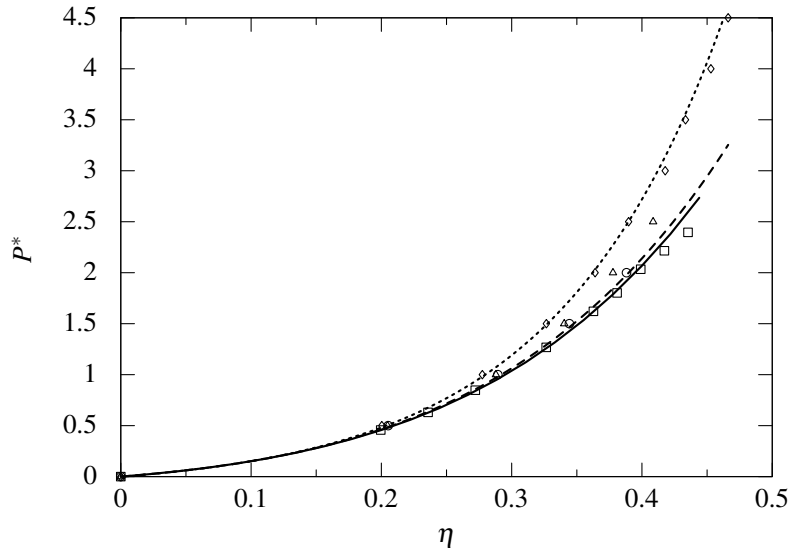}
 \end{center} 
 \caption{\label{fig:eos2} Equations of state giving the pressure from Shinomoto's original route and the virial pressure as functions of packing fraction $\eta$ for the nematic phase of parallel SPCs with different length-to-width ratios.
 The solid and dashed curves give the virial pressure $P^{*}_{\rmv}$ for SPCs with $L/D=\infty$ and $0$ (i.e., hard spheres), respectively.
 The dotted curve represents the pressure $P^{*}_{\rmS}$ for SPCs with $L/D=0.25$.
 The squares, circles, triangles, and diamonds show the simulation results for SPCs with $L/D=\infty$, $2$, $1$, and $0.25$, respectively.
 These simulation data are taken from Ref.~\cite{Veerman1991} for $L/D=\infty$ and from Ref.~\cite{Stroobants1987} for the others.}
\end{figure}

The dependence of $P^{*}_{\rmv}$ on $L/D$ for parallel SPCs is quite weak.
The virial pressure with $L/D = 0$ (i.e., for HSs) and $L/D = \infty$ is shown in Fig.~\ref{fig:eos2}.
For $L/D=0$, the exact expression for the virial pressure becomes
\begin{equation}
 P^{*}_{\rmv} = \eta + 4\eta^{2}e^{5\eta/2}.
\end{equation}
The difference in virial pressures between SPCs with $L/D=0$ and $\infty$ is a few percent at $\eta=0.444$, which is the packing fraction of the nematic-smectic phase transition for SPCs with $L/D=\infty$~\cite{Veerman1991}.
All virial pressures for SPCs with finite $L/D$ are between these two curves (see solid and dashed curves in Fig.\ref{fig:eos2}).

The simulation data in Fig.~\ref{fig:eos2} are taken from the molecular dynamics simulations for $N=1080$ SPCs with $L/D=\infty$~\cite{Veerman1991}, and MC simulations for $N=90$ SPCs with $L/D=2, 1$ and $0.25$~\cite{Stroobants1987}.
The weak dependence of the pressure on $L/D$ is also found in numerical simulations if $L/D \ge 2$; thus, $P^{*}_{\rmv}$ agrees well with these data from $L/D=2$ to $\infty$.
For SPCs with $L/D < 2$, however, the pressure at a given $\eta$ increases as $L$ decreases, and the agreement between the virial pressure $P^{*}_{\rmv}$ and the numerical results worsens.
As expected from the success of Shinomoto's method for HSs, the pressure $P^{*}_{\rmS}$ gives better results for SPCs with small $L/D$; but its validity is limited to $0 \leq L/D \leq 0.25$.

\section{Discussion}
The failure of $g_{\rmc}(z)$ at high densities indicates that the agreement between virial pressures for SPCs from $L/D=2$ to $L/D=\infty$ and those of MC simulations is due to a coincidental cancellation of errors.
Furthermore, the virial method for calculating pressure is not valid for SPCs with small $L/D$, so we are obliged to use Shinomoto's original method; however, there is no theoretical justification to use two different methods depending on $L/D$.
Yet, despite these flaws in the theory, it is interesting that $P^{*}_{\rmv}$ agrees well with the numerical data over a broad range of $L/D$, and that $P^{*}_{\rmS}$ for $L/D = 0$ and $P^{*}_{\rmv}$ for $L/D = \infty$ seem to give the upper and lower bounds for pressures for all $L/D$.

In principle, Shinomoto's method can be generalized to a system of SPCs with arbitrary direction and even to arbitrary hard-core systems.
However, it is difficult to implement this method for such systems.
For example, in many cases, the excluded volume has a complex shape, so calculating the overlap volume of $V_{\text{exc}}(\Omega, \Omega'')$ and $V_{\text{exc}}(\Omega'', \Omega')$ is almost impossible (where $\Omega$ is the orientation of a nonspherical molecule and $V_{\text{exc}}(\Omega, \Omega')$ is the excluded volume of two molecules with orientations $\Omega$ and $\Omega'$).
Numerical estimation of the overlap volume might be possible by MC integration, as done in calculating the virial coefficients of nonspherical molecules.
If such a numerical estimation is performed, Shinomoto's method, which is free from divergence, would have an advantage to the virial expansion, which has a considerably small radius of convergence for strongly anisotropic molecules~\cite{You2005}.

There is no explanation for the insensitivity of the equation of state to the SPC length in $2 \le L/D \le \infty$.
If the system is a fluid of parallel hard ellipsoids, the dimensionless pressure is independent of the molecular length since the system can be mapped to an HS fluid with a scale transformation.
However, the SPC fluid cannot be mapped onto a single system and the above discussion for ellipsoids is not applicable.

\section{Conclusion}
The pair distribution function for an SPC fluid derived by Shinomoto's method agrees well with the result of MC simulations at low density but the agreement worsens at high densities.
From this Shinomoto's pair distribution function, the equation of state is derived by two different routes: Shinomoto's original route and the virial route.
Despite the failure of Shinomoto's pair distribution function, the virial pressure $P^{*}_{\rmv}$ agrees very well with the numerical data for $2 \le L/D \le \infty$ throughout the nematic phase.
The pressure from Shinomoto's original route, $P^{*}_{\rmS}$, agrees well with the numerical data for $0 \le L/D \le 0.25$.
$P^{*}_{\rmS}$ for $L/D=0$ and $P^{*}_{\rmv}$ for $L/D \to \infty$ seem to give the upper and lower bounds for pressures for all $L/D$.






\end{document}